# Dynamic Prize Linked Savings: Maximizing Savings and Managing Risk

Oisin Connolly, Lehigh U

**Abstract:** Prize linked savings accounts provide a return in the form of randomly chosen accounts receiving large cash prizes, in lieu of a guaranteed and uniform interest rate. This model became legal for American national banks upon bipartisan passage of the American Savings Promotion Act in December 2014, and many states have deregulated this option for state chartered banks and credit unions in recent years. Prize linked savings programs have unique appeal and proven societal benefits, but the product is still not available to the vast majority of Americans. There is demonstrated interest in these products, but the supply side may be the bottleneck, because the prevailing consensus is that prize linked savings primarily appeal to low income consumers. This paper examines a less common, dynamic prize, model of prize linked savings and shows why it might result in a larger average account size. The paper proposes three methods of managing risk under this model, and tests two of them using a Monte Carlo simulation. We conclude that both tested methods are effective at mitigating the most severe risks.

1.     Introduction

Prize Linked Savings (PLS) may be new in the United States, but the first PLS program was designed in 1694 and these products have been offered in over 20 countries. Several consistent findings help understanding the appeal and forecasting the reception of this product in the United States.

The UK's Premium Bonds program provides the most striking example of the potential opportunity. Around a third of UK citizens own Premium Bonds and collectively hold over £65 billion in them. (Lobe, Hölzl, 2007) and (Tufano, 2007) studied decades of Premium Bond sales, and both separately found evidence that investor's treat Premium Bonds like a gambling and investment hybrid. Typical of gambling products, the size of the largest prize and skewness of the distribution significantly influenced sales. Sales also responded to the expected return and the return relative to fixed income and equities, like an investment.

    Much of the previous literature also focuses on the unique appeal and the resulting positive benefits of PLS. These programs consistently entice people to save more, act as a more prudent substitute good for gambling and the lottery, and draw unbanked people into the financial system. Guillen and Tschoegl (2002) studied Latin American programs and found that low-income populations and unbanked people expressed higher interest in products like these. Maynard et al (2008) conducted a survey of low income Americans and found that 58% expressed positive interest in prize linked savings. Furthermore, they found that optimistic people, lottery players, non-savers, and the unbanked all expressed higher than average levels of interest. Cole et al. (2014) studied the Million a Month program at First National Bank in South Africa and found that it drew unbanked people into the banking system. Save to Win is the largest PLS program in America, which offers its product through 63 credit unions in 6 states. In their 2015 report the D2D fund reported that 49-57% of surveyed new accountholders identified as non-savers.

    Tversky and Kahneman (1992) demonstrated that people maximize utility as measured by heuristics when making decisions under risk. This framework helps explain what can make PLS so enticing and Pfiffelmann(2007) used prospect theory in trying to optimize the utility of PLS. Individuals are generally more than twice as sensitive to losses than to gains of an equal magnitude. When considered as an investment, this enhances PLS' appeal because people are especially averse to alternative

investments that present the possibility of a loss. Prospect theory also shows that probabilities are considered through a weighing function, which overestimates the chance of low probability events occurring and underestimates the chance of average events occurring. People overestimate their chances of winning, and by a greater degree for smaller odds, which is why skewness improves the perceived valued of PLS and why people undertake seemingly irrational behavior like gambling.

This reduces the uncertainty surrounding the appeal, target demographics, and potential size of PLS in the United States. Yet, no large or national banks have offered these options, and the vast majority of Americans cannot get one of these accounts even if they wanted to. According to the FDIC, there are 9 million unbanked and 24 million underbanked households in the United States. Additionally, the Federal Reserve reports that 53% of Americans lack the personal savings to cover basic expenses for 3 months. Widespread access to PLS may provide the right incentives to nudge people into changing their savings habits. One factor that may contribute to why no large banks have offered these products is the fear of the average account size being too low to be profitable. The most notable example so far, Save to Win, is organized through the work of a nonprofit. It should be noted that the vast majority of existing programs and most referenced in the literature (Save to Win, Premium Bonds, Million a Month, and the programs offered in Latin America) all use a fixed prize model. Most analysis has been conducted on this model, which may have led to presumptions about the appeal of PLS that hindered its spread.

This paper will examine a different approach to PLS in the hopes of further developments in the area. Section 2 will study the predominant model, fixed prizes, of PLS. Section 3 will study an alternate model, dynamic prizes. Section 4 compares the two models. Section 5 proposes methods of managing the risks inherent the dynamic prize model. Section 6 details the methods used to run a Monte Carlo simulation. Section 7 analyzes the results of these simulations. Section 8 concludes the paper.

## 2. Fixed Prize Model

### 2.1 Definition

The vast majority of existing PLS programs are modeled in a similar way, but this model does not properly incentivize large amounts of saving. These programs choose and advertise fixed prize sizes and frequency (e.g. "One annual prize of $100,000", "Ten daily prizes of $10"). The other feature is that an investor's chances of winning is proportional to the amount in their account. In this way, the expected return is held constant for all investors, even though the actual return will be random. This model is often described as a lottery with guaranteed return of principal, which is an apt comparison because buying two tickets doubles your chance of winning the lottery.

### 2.2 Prospect theory – No expectation of growth

Prospect theory is used below to analyze the utility of the simplest possible fixed prize setup. This setup has one prize and will return the principal if the prize is not won. We also assume a constant change in probability for each additional unit saved. The change in probability will not be exactly constant because the total pool of funds that can be selected to win also grows as the investor deposits more. With a large enough system it will be near enough to constant that this can be ignored.

The following forms the basis of cumulative prospect theory

$$V(X) = V^+(X) + V^-(X) = v^-(x)\,\pi^-(p) + v^+(x)\,\pi^+(p)$$

$$v(x) = \begin{cases} x^\alpha & \text{if } x \geq 0 \quad \text{where } 0 < \alpha < 1 \\ -\lambda(-x)^\beta & \text{if } x < 0 \quad \text{where } 0 < \beta < 1 \end{cases}$$

$$w^+(p) = \frac{p^\gamma}{(p^\gamma + (1-p)^\gamma)^{1/\gamma}}, \quad w^-(p) = \frac{p^\delta}{(p^\delta + (1-p)^\delta)^{1/\delta}}$$

Applying our simplest model of a fixed prize system we find.

$$V^+(X) = v^+(y)\pi^+(Xc)$$

where X is amount in account, y is the prize amount, and c is the probability growth constant

$$V^-(X) = v^-(0)\pi^+(1 - Xc) = 0$$

Using the product rule we find:

$$V'(x) = v'^+(x)\pi^+(p) + v^+(x)\pi'^+(p) + v'^-(x)\pi^-(p) + v^-(x)\pi'^-(p)$$

$$V''(x) = v'^+(x)\pi'^+(p) + v''^+(x)\pi^+(p) + v^+(x)\pi''^+(p) + v'^+(x)\pi'^+(p)$$

$$V''(x) = \quad ( 0 ) \quad + \quad ( 0 ) + ( + * - ) + \quad ( 0 )$$

If the program is of a reasonable size an investor will have a relatively low chance of winning. The corresponding part of the probability weighing graph is concave, so the second derivative with respect to the amount saved will be negative. Given that the value of a prize is a constant, the first and second derivative of the value with respect to the amount in the account will be zero. Therefore, when taking the second derivative relative to the amount saved, only one term remains and it will be the product of a positive and negative number. There is no utility loss if the prize is not won, so the total second derivative with respect to savings will always be negative.

This demonstration can intuitively be explained as well. Under prospect theory, the largest increased in perceived odds comes immediately and the increase in perceived odds for continued savings is continually decreasing. This phenomenon is not likely to encourage people to save large amounts under this design. With Premium Bonds the smallest possible account is £1 and there is a maximum account size of £50,000. While both accounts overestimate their chances of winning, the large account cannot appreciate that they have increased their odds by 50,000 times. The odds are not only too small to meaningfully comprehend, but the exact odds are usually not told to depositors or updated in real time. These odds are hard to calculate as the total amount in deposits is constantly in flux, which means that people do not get the positive reinforcement that they have increased their chances of winning when they save more.

### 2.3 Prospect theory – Expectation of growth

Taking this analysis further, we can account for the fact that before signing up, investors may consider options that grow their assets at a risk-free rate, as well as alternatives that carry varying degrees of risk and generate higher expected return. This framing means that when thinking about an investment, future expectations usually factor in some growth rate. Given these assumptions, not being selected to receive the prize would be viewed as a loss, even though the absolute position is constant. Further, if an account does win, the gain must be the prize less this growth. As the account grows, the absolute gain represented by alternatives increases proportionally.

Applying this model of a fixed prize system we find.

$$V^+(X) = v^+(y-xr)\pi^+(Xc)$$

where X is amount in account, r is the growth rate, y is prize amount, and c is the probability growth constant

$$v(y-Xr) = (y - Xr)^\alpha \qquad v'(y-Xr) = \frac{-\alpha r}{(y-Xr)^{1-\alpha}} \qquad v''(y-Xr) = \frac{(\alpha-1)\alpha r^2}{(y-Xr)^{2-\alpha}} \qquad \text{where } 0 < X < y/r$$

$$V''^+(x) = v'^+(x)\pi'^+(p) + v''^+(x)\pi^+(p) + v^+(x)\pi''^+(p) + v'^+(x)\pi'^+(p)$$

$$V''^+(x) = ( - * + ) + ( - * + ) + ( + * - ) + ( - * + )$$

We can establish that the second derivative of the prize utility with respect to the amount in the account will again always be negative for the defined region of X. Each of the four terms that make it up are the product of a positive and a negative number. As long as Xr < y, the value function will be positive. This is a reasonable assumption, because an investor would be irrational to give up a guaranteed return for a small chance to win a sum that represents less than the assumed growth rate. As the amount in the account increases, the value of a prize is worth less but the change of winning increases, which accounts for both of the first derivatives. The second derivative of the value function will always be negative as long as $0 < \alpha < 1$, which is an assumption of prospect theory because it accounts for diminishing returns on gains and losses. This is the most general prize model and relies on only a few assumptions, but shows that the second derivative of utility gains with respective to savings will be negative for all X.

$$V^-(X) = v^-(-xr)\pi^-(1 - Xc)$$

$$V'^-(X) = v'^-(x)\pi^-(p) + v^-(x)\pi'^-(p)$$

$$V'^-(X) = (\ -\ *\ +) + (\ -\ *\ -\ )$$

$$V''^-(x) = v'^-(x)\ \pi'^-(p)\ +\ v''^-(x)\ \pi^-(p)\ +\ v^-(x)\ \pi''^-(p) + v'^-(x)\ \pi'^-(p)$$

$$V''^-(x) = (\ -\ *\ -\ )\ +\ (\ +\ *\ +\ )\ +\ (\ -\ *\ \ +\ ) + (\ -\ \ *\ \ -\ )$$

The first and second derivative of not winning a prize are a mix of positive and negative terms. The value here will now always be negative, while the probability will still always be positive. Increasing the amount in the account lowers the odds of not winning a prize, but means that the loss will be worse. The corresponding part of the probability weighing curve for high probabilities is convex, which accounts for the second derivative with respect to the amount in the account being positive. The second derivative of the value function is also positive, because the same diminishing effect applies to losses. What we can conclude is that the first derivative with respect to X is very likely to be negative at lower values of X. Consider that at lower account sizes, the two parts of the negative term will be at their highest absolute value, because the odds of not winning are highest and the second derivative of the value function is positive so the first derivative will be at its lowest initially. The utility cost of a loss, one of the terms of the positive term, is at its lowest absolute value initially. The first derivative of the probability weighing function is at the greatest absolute value initially, but assuming a program of a reasonable size the difference in probabilities between someone with no investment and a large account is still relatively small. Therefore, the marginal change is perceived probability is not likely to be excessively large. One would also intuitively expect that the more money an investor had to invest, the more cautious they would be of the possibility of not gaining anything, even if that account is less likely, so the first derivative being negative seems reasonable.

Examining the equations further reveals how all this comes into play to deter large savings accounts. At the point where X = y/r the utility value of the prize will be 0, and putting in additional savings causes a loss from the value function, even when winning the prize. This point corresponds to where the prize is equal to the assumed growth of the account, so it represents an absurdly large account. But for the utility of a prize to become 0, given that the second derivative is always negative, the prize's marginal utility with respect to savings would have to become negative prior to this point. At that point it would be irrational to continue saving. The gain must also compensate for the loss, which we know will always have negative utility. We also believe it is likely that the more one saves, the larger the possibility of losing looms. Realistically, the potential gains must compete with other forms of saving and consumption, so savings are likely to stop before the marginal utility of the prize becomes 0.

This can also be demonstrated intuitively. As an individual dedicates more assets to an account, the opportunity cost becomes greater, but the size of prizes are fixed. For a setup with multiple prizes, one can imagine a situation where winning the smaller prizes represent a utility gain for a small account, but represents a utility loss for a large account. Premium Bonds accounts can win multiple times each month which partially helps address this problem, but it is still an illustrative example. Both a minimum and maximum size account have the chance to win prizes ranging from £25 to £1,000,000 every month, and an expected return of 1.25%. Accounts that hold the maximum, £50,000, are virtually guaranteed to win something each year, but can expect to receive less than £500 over the course of a year 29% of the time. A substantial part of the return effectively minimizes losses for larger accounts, but any win represents a gain for the smallest accounts. Gains of tens of thousands are conceivable if one had put the maximum account into equities, but there are no alternatives that offer the possibility of a comparable return for a minimum account. The opportunity cost is another factor that is problematic if one is trying to encourage significant savings.

## 2.4  Evidence in the Literature

Diminishing returns under this model have not been intentionally studied, but evidence of this can be found in empirical research. Cole et al. (2014) did the regression in Figure 1 to control for age, gender, race, and scaled relative to income to compare FNB employees who opened a Million a Month account to those who did not. They wrote,

> About 2 months prior to opening a PLS account, total net savings begins to increase, with a large jump in savings occurring on the month that the MaMa account is opened. From this point onwards, MaMa participants maintain roughly 1% of annual income more in total net savings at the bank, relative to non-participants.

In the two years after a participant opens an account, net savings does not continue to increase significantly after the initial investment. The next regression examines the standard savings accounts of those that opened a Mama account relative to the control group. Figure 2 shows that the participant's standard savings accounts gradually increases relative to the control group throughout the following two years. This is a more interesting finding, because it represents savings that participants chose not to put into their MaMa accounts and shows evidence that the equilibrium point for PLS was quickly reached.

Maynard (2007) prepared a report for D2D which contained Figure 3. It shows the average deposit per account at Centra Credit Union on a weekly basis after PLS was first offered there. Accounts are being rapidly opened during the first few months, but from mid April to the end of July there were always around 1300 accounts and 0 accounts were opened in many of the later weeks. The average started at around $220 in mid January and increased over the next few months, but eventually settled at around $375. The author acknowledges that "most customers were maintaining their deposit balances in the product following the first three months of the launch". This trend is also seen in a Save to Win report on Washington State in 2014. The total deposits of accounts from the previous year do not dramatically change throughout the year, which can be seen in Figure 4. Both examples show that people tend to quickly stop depositing more rather than continually build on their savings.

Save to Win awards tickets for each $25 in deposits, but caps each account at 10 tickets per month to encourage the habit of consistent deposits. A report on Save to Win in Michigan (2011) noted that,

> The savings pattern of new participants in 2009 and 2010 were very similar—while balances grew throughout the year, median deposit activity in the last few months trended towards zero … Almost 15% of accountholders earned only one (1) raffle ticket … many of them may only be interested in earning one chance to win the $100,000 grand prize…. While 9% of accountholders have more than $3000 in savings, only 1% earned the maximum 120 entries. Many of these high balance earners are participating but not motivated to save regularly to earn their maximum entries.

This shows evidence that under this model people will be inclined only to save the minimum allowed investment, which is consistent with the theory that the lowest odds are overestimated to the greatest degree. Additionally, 8 out of 9 depositors who deposited over $3000 did not structure their deposits strategically, which indicates that it is likely not a lack of funds that cause people to stop saving.

All these empirical examples show that while many people are enticed by PLS enough to open accounts and save initially, increasing their odds does not motivate them to continue saving.

## 3. Dynamic Prize Model

### 3.1 Definition

The dynamic prize model chooses prizes that are a percentage of the amount in the account (e.g. "the winner will receive 100% interest). All accounts have an equal chance of winning a given prize, but rather than the odds of winning being adjusted, the prize amount changes as investors save more. The return still stays constant for all accounts. This model is more analogous to investing in equities, because the amount invested does not affect the probabilities of an outcome materializing, but does change the absolute gain that outcomes represent. This model is better designed to correct for two of the factors that cause diminishing marginal utility under the fixed prize model and result in many small accounts. Probability weighting does not come into play as one increases their account size, because the probability and perceived probability of winning are not dependent on account size. Even when one factors in the opportunity cost, the size and utility of a prize will only increase as the account size increases.

### 3.2 Prospect theory – Expectation of growth

$V^+(X) = v^+(x(w-r))\pi^+(p)$      where p is the odds of winning, x is the amount in account, w is the size of the prize, and r is the growth rate

$V'^+(x) = v'^+(x)\pi^+(p) + v^+(x)\pi'^+(p) = \frac{\alpha(w-r)}{x^{1-\alpha}} \pi^+(p)$

$V''^+(x) = v'^+(x)\pi'^+(p) + v''^+(x)\pi^+(p) + v^+(x)\pi''^+(p) + v'^+(x)\pi'^+(p) = \frac{\alpha(\alpha-1)(w-r)}{x^{2-\alpha}} \pi^+(p)$

Any terms featuring a derivative of the weighing function become zero, because the probability is a constant. The last line shows that this model also has diminishing marginal utility, because $\alpha - 1$ will be negative. However, the reason for the diminishing marginal utility here is that the second derivative of the value function is negative. Each additional dollar won is worth a little less utility, but this is an inherent property of accumulating wealth that is shared by any method of saving. Unlike the fixed prize model, it is not possible for the marginal utility of a win to become negative as the account grows.

Every time an investor deposits additional funds the potential prize grows bigger, which may be easier to appreciate than the difference between two extremely low probabilities and is certainly easier to display in real time. Previous studies have shown there is utility from the excitement and anticipation of large prizes. Under the fixed prize model many investors deposit the bare minimum to be eligible for the largest prize, but the smallest deposits results in the least exciting prizes under the dynamic prize model. This model encourages investors to continually save more so that their prize will keep getting bigger and more exciting. This design may mean that investors are likely to save more on average.

## 4. Comparing Models & Implications

The fixed prize model allows an organization to know their costs with certainty. The comparative ease of managing a fixed prize model may account for its proliferation. However, we have found that there are two phenomena under the fixed prize model that create a diminishing marginal benefit to saving. The consensus has become that PLS has much greater appeal among poor consumers and the unbanked, but this could be due to the prevalence of the fixed prize model. The fixed prize model is so prevalent that some literature uses lottery linked deposit accounts as synonymous with PLS. It has also been observed that the lottery is strongly regressive; income has a negative correlation with the percentage of income spent on the lottery. For poorer consumers, the lottery increasingly represents the only way of accumulating significant wealth with a minimal investment, so the fixed prize model offers a preferable alternative. Meanwhile, random returns, such as those found in equities, and gambling have broad appeal. The perception of PLS may apply more to the fixed prize model, than randomized returns in general, and this is an area for future study.

The dynamic prize model may encourage higher average savings. Everyone overestimates the odds of low probability events, but optimistic people and gamblers are still good target customers for this more exciting method of saving. Poor consumers and the unbanked will still find this more attractive than a traditional savings account. It still offers the potential for significant return with only a minimal investment and no possibility of a loss. Unlike the fixed prize model, people are probably not as likely to stop increasing their savings after reaching the minimum account level.

Many of the parties involved have an incentive to encourage higher levels of saving. Policy makers in the United States deregulated PLS with the hopes of encouraging saving and bringing some of the millions of unbanked into the banking system. The private sector needs a minimum level of saving to ensure viability. Maintaining an account carries some fixed cost, so larger accounts are more profitable. Lastly, investors are better off if they save more, especially those who otherwise would not be saving or would spend the funds gambling.

The disadvantage of the dynamic model is the risk for the supply side. Organizations cannot forecast the cost of prizes with certainty, which has problematic implications. MMax is a savings account offered by a French insurance company, les Mutuelles du Mans, and is the only example of this model known to the author. Pfiffelman looked at MMax and found that they offered a guaranteed 2.5% interest, plus chances to win 5%, 10%, and 20% prizes. Pfiffelman optimized the prizes offered by the program using cumulative prospect theory. The first test minimized the expected cost while staying at or above a given level of utility. Then the consumer's utility was maximized while staying at or below a given level of expected cost. In both cases, the guaranteed interest and first two prizes were set as low as the constraints would allow and the size of the last prize was maximized.

This example illustrates the inherent tradeoff between risk and appeal when choosing the prize sizes under the dynamic prize model. The risk under this model stems from the fact that there will be accounts with varying amounts in them. The cost of a given prize will be highest if the largest account is chosen, and will be minimized if the smallest account is chosen. MMax chose to reduce risk by offering relatively small prizes to reduce the range between the largest and smallest possible cost. However, the research on PLS indicates that large top prizes and a high skew maximizes the appeal of PLS, which is reflected in Pfiffelman's optimization. The risk and its tradeoff with appeal presents a significant roadblock to this model. The remainder of the paper will examine approaches that mitigate the disadvantages of the dynamic prize model.

## 5. Methods of Reducing Risk

### 5.1 Understanding of Dynamic Prize Risks

First, we examine a generic example of a dynamic prize model from the supplier's perspective. The expected cost of a prize will always be equal to the mean account size times the percentage of that prize, because all accounts have the same chance of being chosen. The effective interest paid out for the entire program will be the total cost of all prizes over a given time divided by the total assets being managed. It follows that the expected interest would be the sum of all prize percentages divided by the number of accounts.

While the expected costs and interest can be easily derived, the variance depends on the specifics of the situation. Suppliers can handle some level of variance in cost, because this is true of most expenses. The risk can become untenable in two ways. If there is a significant risk of the costs being higher than revenue, then the profitability of this model may be too unreliable to be worth the undertaking. Also, if there is a risk of a single prize being significant relative to the size of the firm, this could create immediate liquidity problems. Setting conservative prizes helps, because moving the expected cost farther from the highest acceptable cost level will reduce the chance of exceeding it. This could be achieved by adjusting the frequency and number of prizes without compromising the ability to offer large prizes. A risk management method that does not lower the expected return would be preferable from a consumer's perspective. Three methods of doing so are proposed below that may be used separately or in conjunction.

**5.2   Bracketing**

If multiple prizes of the same size are being distributed, one approach would be to order the accounts by size. Then break the program into n brackets, where n is the number of prizes to be given out, so that each bracket has the same number of accounts. Next, randomly choose one account from each bracket. This eliminates the possibility that the n biggest accounts all win at the same time, but maintains the condition that all accounts have an equal chance of winning a given size prize. The disadvantage is that this would raise the minimum possible payout, but this is much less of a concern.

**5.3   Caps**

The institution could place a limit on the maximum that any one account can hold. Without a limit on the size of accounts, there is no limit to the cost of prizes. This is especially problematic when offering very large prizes. The drawback of this approach is that the largest accounts hold a disproportionate amount of the total deposits and the intention of this model was to raise the average account size.

**5.4   Insurance**

Insurance or customized financial instruments could reduce the variability of costs by paying out under certain conditions. Protecting against high costs will likely result in a premium above the expected cost of the prizes. This could be offset by using a box spread approach and agreeing to pay an outside party when the prize costs are low. These instruments have a return that is random and will not be correlated to any asset class, which is an attractive selling point. This may represent the only method of eliminating any variance in cost.

**6.   Monte Carlo Methodology**

Bracketing will increase some payouts by guaranteeing that an account in the largest bracket is chosen and caps will lower the average account size. Using reasonable assumptions and Monte Carlo methodology, we tested how these methods preform as risk management tools. A Pareto distribution will be used as a starting point to model what the distribution of account sizes might look like. This distribution works well for skewed distribution with long tails, which is why it is frequently chosen to model income and wealth.

Pareto Probability Distribution: $f(x) = \frac{\alpha b^\alpha}{x^{\alpha+1}}$

Pareto Distribution Mean: $E(X) = \frac{\alpha b}{\alpha - 1}$

Pareto to Gini Coefficient Conversion: $G = \frac{1}{2\alpha - 1}$

Pareto Quintile Function: $F^{-1}(p) = \frac{b}{(1-F)^{1/\alpha}}$

The two parameters for this distribution, $\alpha$ and b, control the shape and scale respectively. The scale parameter is equivalent to the minimum possible value under the distribution. Two sets of values were chosen to account for possible variations of the distribution, (1.12, 250) and (1.04, 150). The Gini coefficient, given as G, is often used as a measure of inequality. A 2015 Allianz report found that the Gini wealth coefficient in the United States is 0.8065, which allows us to estimate the α parameter as 1.12 using the above formula. The other set of parameters uses a lower scale parameter to simulate a more unequal distribution, which presents more of challenge from a risk management perspective. The average account size of Premium Bonds was around £2,350 in July 2014 a month after the maximum was raised from £30,000 to £40,000. Using the shape parameter, the minimum value of 250 was chosen to give a comparable average account size. The other minimum value can be thought of as more attainable minimum. We are using an uncapped distribution and we expect the dynamic model to result in higher average savings, so an expected value of 3900 is also reasonable. All the payments will be scaled relative to the expected payment, so the absolute size of the accounts is not consequential.

As a measure of risk we used Value at Risk, because we are most concerned with low probability events that represent multiple large accounts being chosen. The quantile function is used in Table 1 to show what the distribution looks like. The median account value will be a fraction of the average account size and less than a tenth of accounts exceed the average account size. The small tails will be significantly more than the average account size, which represents a significant risk if chosen.

The simulations model a program with 100,000 accounts. All the prize offerings result in an expected interest rate near 1%. The prizes were chosen to be large enough to be interesting and can be easily explained in marketing material. The four options are 1,000 double your money prizes (100% prize), 500 triple your money prize (200% prize), 100 "add a zero prizes" (900% prize), and 10 "add two zeros prize" (9900% prize).

To test the effect of bracketing each run involved randomly generating 100,000 numbers from the given Pareto distribution to represent accounts. For all four prize options, the respective number of accounts were randomly chosen 10,000 times to represent winners. All the chosen accounts were summed and multiplied by the respective prize multiple to represent the bank's total payout. Then all 10,000 payouts were ordered from least to greatest. The 9500$^{th}$, 9900$^{th}$ and 9990$^{th}$, and largest were selected as approximations of 5%, 1%, 0.1%, and 0.01% VAR. Lastly, the worst possible payout given that sample of accounts was calculated. All payouts were divided by the expected prize payout, which was found using the mean of the sample of 100,000 Pareto values. The process of selecting prizes was then repeated as detailed above, but using the bracketing methodology so the results could be compared. All of the above constitutes one run and 200 runs were performed for each set of parameters.

To test the effect of caps each run involved randomly generating 100,000 numbers from the given Pareto distribution. For each prize option, the respective number of accounts were randomly chosen 1,000 times, summed, and multiplied by the appropriate prize percentage to represent payouts. After being ordered by size, the 950$^{th}$, 990$^{th}$ and largest payouts were selected as approximations of 5%, 1%, and 0.1% VAR and the worst possible payout was calculated. To test the effect of a cap any account larger than 250,000, the FDIC coverage limit, was reduced to this new maximum value and the mean of the whole sample was recalculated. To find the new prize payouts the same accounts as before were reduced if they were larger than the cap and summed. The payouts were resorted by size at this point, so the order may change at this point. The cap process was repeated with 50,000, the current Premium Bond limit, and 10,000 as maximum caps. The payout was divided by the expected prize payout, each time using the mean of the entire sample after the cap change. All the above constitutes one run and 2000 runs were performed for each set of parameters. More runs were conducted because this data may be more sensitive to the specifics of the mean of the sample, how the accounts are generated in relation to the caps, and the distribution of chosen accounts in relation to the caps. Each prize option was only chosen 1000 times in each run so that the simulation was less resource intensive.

The Monte Carlo simulations were run using Python. The code and output can be found at https://github.com/oconnolly/Prize_Linked_Savings

## 7. Results

### 7.1 Bracketing

In Table 2, for each set of parameters and prize setup, the first two columns of data represent the average of the 200 payouts after each payout was scaled. The payouts were scaled so the expected payout for that sample of program values is equal to 1, so higher is worse from the supplier's perspective.

As expected, the payouts tended to be higher at more extreme VAR levels and with the more skewed distribution. The worst payouts will tend to be comprised of accounts from the tail of the distribution, so a more skewed distribution increases this potential. The last column represents a different perspective by showing the percentage of times when the payout was higher after bracketing.

The Monte Carlo simulation showed that bracketing is an effective method of risk management. The highest possible payout average when bracketing was significantly lower than the worst possible average without restrictions. For both sets of parameters with the 100% and 200% prizes, the worst possible prize average after bracketing was lower than the 0.1% VAR average without bracketing. The data shows that bracketing helps mitigate less extreme risks as well. Table 3 displays the percentage difference between the average bracketing and random prize payouts. It seems that the more bracketing was used the more it helped; the last two rows highlight that choosing more prizes tended to result in a greater reduction to the VAR approximations and the greatest possible payout. The bracketing average is lower in 28 of the 32 direct VAR comparisons. In 11 of the 32 VAR comparisons, none of the 200 payouts were higher after bracketing. Overall, bracketing resulted in a higher payout in only 14.4% of the VAR comparisons, 1052 of the 6400 comparisons.

The outliers reveal important insights about bracketing and the number of prizes offered. Table 4 shows that the larger prizes have a higher standard deviation. Holding everything else constant, the size of prizes does not affect the variance, because everything is scaled to the expected payout. The payouts, percentage of prizes, and expected payout all are directly proportionally. Therefore, the higher variance is a result of choosing fewer prizes. For any real application, when choosing prizes without restriction there would be a finite mean and variance, so the central limit theorem confirms this trend in the random case. In the case of bracketing, awarding fewer prizes means that each individual prize can be chosen from a wider range so this also makes sense.

The relationship between bracketing and the risk level is more complicated. Bracketing seems to be very effective at the highest and lowest VAR levels examined, but less so for the middle levels. In 6 of the 8 0.01% VAR comparisons the bracketing payment was lower in all 200 cases. For 5 of the 8 prize and parameter setups, the highest percentage of bracketing payouts increased at the 1% VAR level, and the 0.1% VAR has the highest percentage in the other 3. The lowest average percentage is at the 5% VAR level, where only 4% of the bracketing payouts were higher.

The interplay of two trends may contribute to this finding. The gap between the average payout when choosing without restriction and the bracketing payouts is the largest at the most extreme VAR levels. Bracketing reduces the worst possible payouts by the largest degree as seen in the last line of Table 3. Bracketing also result in a guaranteed increase in the minimum possible payout and for similar reasons one would expect it to be a dramatic increase. It is harder to ascertain if this trend line holds for probabilities in between. In Table 3 for the 100% and 200% prizes at both sets of parameters, the decrease from bracketing at the 0.01% VAR level is significantly more than the other VAR levels. Due to the most prizes being chosen these are the least likely to deviate from their true value, so this supports the theory that bracketing creates larger average decreases in payouts at higher risk levels. On the other hand, the testing methodology may contribute to more variability at more extreme VAR levels. Table 4 shows that the more extreme the VAR approximation the larger the standard deviation of the payouts. The 5% VAR approximations are less than 500 of the 10,000 payouts for that prize and parameter setup, whereas the largest VAR approximation is, by definition, an outlier. The less extreme VAR values may be less likely to deviate significantly from the true VAR level. The large separation in true average payouts means that the increased standard deviation does as much of a role at the highest VAR levels. At lower VAR levels even though the reduction in risk is less, the decreased standard deviation of the VAR approximations makes bracketing seem more effective.

The areas where bracketing was not as effective generally occurred where we would expect them to. Three of the exceptions with a higher average are from the 0.1% VAR comparisons and one is from the 0.01% VAR comparison. Three of the exceptions with a higher average are when the least prizes are chosen and the last is from the second least. Both comparisons with a percentage over 40% are from the largest prize and the 0.1% VAR level. These conditions all result in a high standard deviation and a gap between the true averages that is not too large. We conclude that the exceptions are likely due to these factors rather than bracketing not working under these circumstances. Further testing should be done on if the bracketing average would converge to a lower level than the random average as the number of runs approached infinity.

### 7.2   Caps

In Table 5, the columns under cap amounts represent the average payout across the 2000 simulations relative to the expected interest. Lowering the cap changes the expected interest and each simulation tests the same accounts at all four cap levels. Lowering the cap cannot increase the total prize payout, but it can increase the payment relative to the expected interest. This provides a look at a cap's effectiveness as a risk management measure with the acknowledged tradeoff of lowering the average account size. The percentage columns represent cases where the payout average relative to the expected payout was higher compared to the previous cap (e.g. 250000 is being compared to the uncapped payouts, 10000 is being compared to the 50000 payouts). For the $\alpha = 1.04$, $b = 150$ distribution, 1.268%, 0.238%, and 0.045% of values will be larger than each of the respective caps. For the $\alpha = 1.12$, $b = 250$ distribution, 1.606%, 0.264%, and 0.044% of values will be larger than each of the respective caps.

The Monte Carlo simulation showed that caps are generally an effective method of risk management. The worst possible payout relative to the expected payout never increased in 48,000 comparisons, which encompasses three cap levels, two sets of parameters, and four prize setups. The payment increased 8.92% of the time for the 0.1% VAR approximations, whereas it increased 26.56% of the time for the 5% VAR approximations. The caps are better at managing more severe risks, because they consist of larger accounts being chosen. The reduction in payouts needs to offset the lower expected payout for the ratio to decrease. Scenarios that consist entirely of accounts below the cap will always be worse off, because there will be no relief from the cap to compensate for the lower average account size.

The average payout was lower in 62 of the 72 VAR comparisons, but breaking it out by cap size and prize setup reveals stark differences and important trends. 8 of the 10 increased average ratios were from adding the 250,000 cap. The two other increases were from decreasing the cap from 250,000 to 50,000 at the 9900% prize level. 8 of the 10 increases are from the 9900% prize setup. The two remaining increases are from the 900% prize level when adding a 250,000 cap. The payout ratio increased in only 1.46% of the 72,000 comparisons with the 100% and 200% prize setup, but 28.19% of the 72,000 comparison with the 900% and 9900% prize setup. To explain why, consider that under the distributions, 0.011% and 0.009% of generated values should be larger than 1,000,000. These accounts are unlikely to be chosen, even for relatively high risk approximations, when only choosing 100 or 10 prizes. Yet, just one of these accounts being reduced to 250,000 will reduce the average account size for the entire program by at least 7.5. When using higher caps and choosing fewer prizes, it is less likely that an account that significantly exceeds the cap will be selected for a prize, so the decrease in payouts is less likely to keep pace with the decrease in the average account size.

Using lower caps and choosing more prizes causes the benefits to materialize more consistently. For 23 of the 24 VAR comparisons the 10,000 cap had the lowest average payout ratio. The exception was at the 5% VAR level with 9900% prizes, which are both conditions that we believe make caps less likely to effective. Adding a 250,000 cap increased the payout ratio in 36.05% of the comparisons, while reducing the cap from 50,000 to 10,000 did not increase the payout ratio in any of the 48,000 comparisons. Reducing the cap from 250,000 to 50,000 did not increase the payout ratio for any of the 36,000 comparisons with 100%, 200% or 900% prizes.

### 8.   Conclusion

We presented an analysis using prospect theory showing why the design of the predominant model of prize linked savings causes saving to have diminishing returns and cited empirical examples in the literature that corroborate that theory. We suggested that an alternate, dynamic prize, model may be better suited to encouraging a high level of saving. Using the premise that this would make it more attractive to suppliers and enable the proliferation of prize linked savings in the United States, we proposed four strategies to reduce the risk associated with the dynamic prize model. We conducted a Monte Carlo analysis to test bracketing and caps, and found both to be effective means of managing risks under the dynamic prize model. Empirical research should be done on the dynamic prize model and its appeal as compared to the fixed prize model. Pricing of securities that would eliminate the variability of payout costs is another important area of future research.

**Figure 1**

Cole et al (2014) page 43: http://www.kellogg.northwestern.edu/faculty/iverson/papers/MaMa.pdf

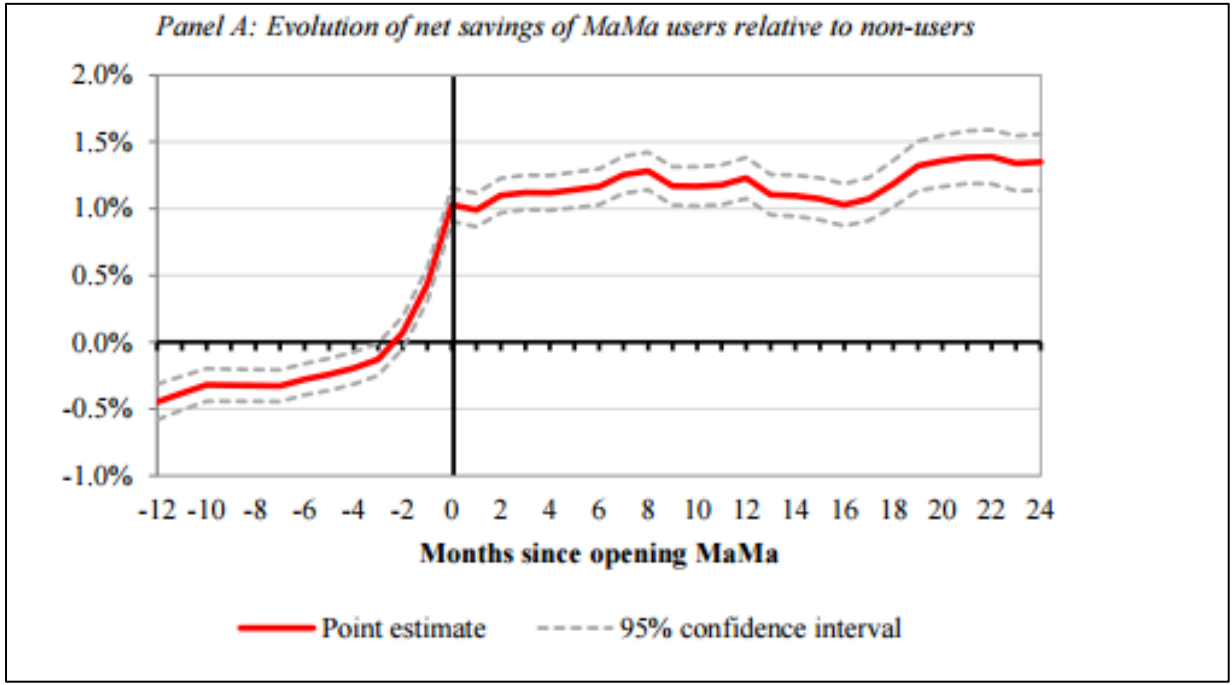

**Figure 2**

Cole et al (2014) page 43: http://www.kellogg.northwestern.edu/faculty/iverson/papers/MaMa.pdf

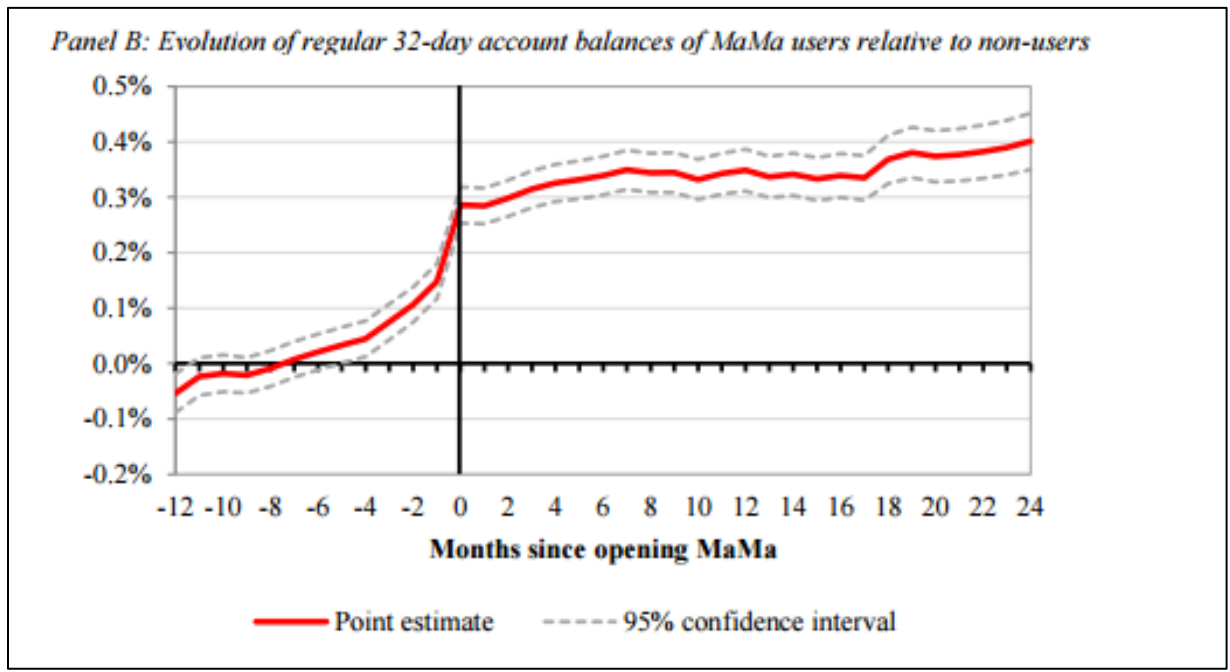

**Figure 3**

Maynard (2008) page 14: http://buildcommonwealth.org/assets/downloads/prize-based-savings.pdf

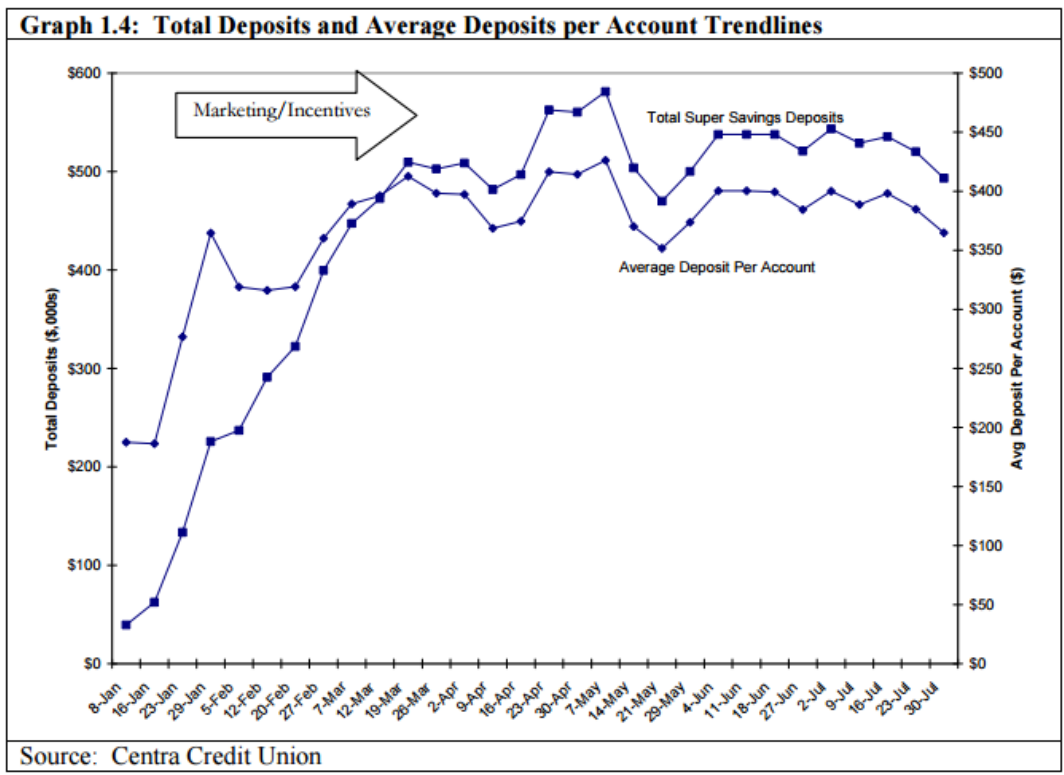

**Figure 4**

Washington D2D Fund (2014) http://buildcommonwealth.org/assets/downloads/STW_WA_2014.pdf

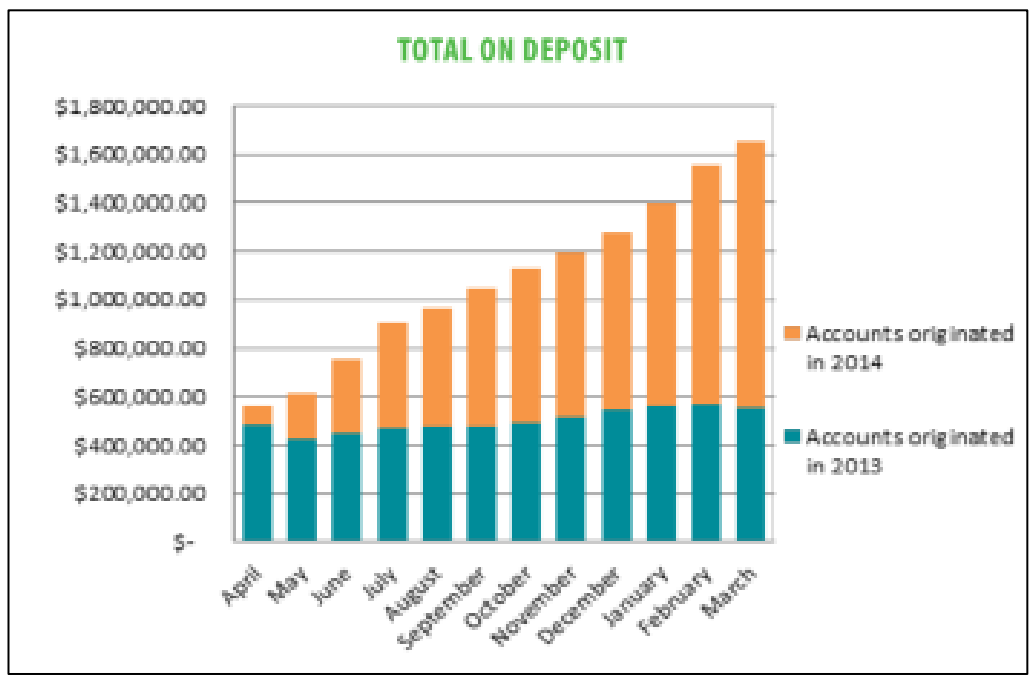

## Table 1

This table give an understanding of the Pareto distribution for each set of parameters.

| Parameters ($\alpha$,b) | E(X) | Median | 90th Percentile | 95th Percentile | 99th Percentile | 99.9th Percentile | 99.99th Percentile |
|---|---|---|---|---|---|---|---|
| 1.04, 150 | 3900 | 292.11 | 1372.87 | 2673.51 | 12565.16 | 115002.33 | 1052555.74 |
| 1.12, 250 | 2333 | 464.21 | 1953.43 | 3627.22 | 15263.51 | 119264.56 | 931898.43 |

## Table 2

First two columns under the parameters represent the average of 200 payouts relative to the expected payout for that sample. The percentages represent how many of the bracketing payouts were higher.

### 1000 100% Prize

| | $\alpha$ = 1.04, b = 150 | | | $\alpha$ = 1.12, b = 250 | | |
|---|---|---|---|---|---|---|
| **Risk** | **Random** | **Bracket** | **Percentage** | **Random** | **Bracket** | **Percentage** |
| **5%** | 2.044715 | 1.864649 | 10.5% | 1.815244 | 1.672558 | 9.5% |
| **1%** | 11.13281 | 9.257685 | 40.0% | 8.092922 | 7.448865 | 35.5% |
| **0.1%** | 15.23429 | 14.40369 | 0.0% | 11.79225 | 11.12558 | 0.0% |
| **0.01%** | 18.56896 | 14.41345 | 0.0% | 14.2899 | 11.13466 | 0.0% |
| **Worst** | 60.90616 | 14.42867 | 0.0% | 51.56385 | 11.14881 | 0.0% |

### 500 200% Prizes

| | $\alpha$ = 1.04, b = 150 | | | $\alpha$ = 1.12, b = 250 | | |
|---|---|---|---|---|---|---|
| **Risk** | **Random** | **Bracket** | **Percentage** | **Random** | **Bracket** | **Percentage** |
| **5%** | 1.998953 | 1.816534 | 6.0% | 1.822109 | 1.657789 | 3.0% |
| **1%** | 7.678657 | 7.130647 | 31.0% | 5.962076 | 5.623056 | 27.0% |
| **0.1%** | 28.63217 | 28.17555 | 0.0% | 21.93715 | 21.5374 | 0.0% |
| **0.01%** | 32.69455 | 28.19015 | 0.0% | 25.02065 | 21.55016 | 0.0% |
| **Worst** | 111.1815 | 28.20754 | 0.0% | 92.22351 | 21.56764 | 0.0% |

### 100 900% Prizes

| | $\alpha$ = 1.04, b = 150 | | | $\alpha$ = 1.12, b = 250 | | |
|---|---|---|---|---|---|---|
| **Risk** | **Random** | **Bracket** | **Percentage** | **Random** | **Bracket** | **Percentage** |
| **5%** | 1.966473 | 1.750604 | 0.5% | 1.88734 | 1.683557 | 0.0% |
| **1%** | 7.19334 | 6.852016 | 26.0% | 5.973321 | 5.779805 | 29.0% |
| **0.1%** | 74.59233 | 83.57696 | 36.5% | 61.36636 | 58.40221 | 28.5% |
| **0.01%** | 141.8587 | 138.6806 | 0.0% | 107.9161 | 105.122 | 0.0% |
| **Worst** | 438.479 | 138.7066 | 0.0% | 349.0005 | 105.1494 | 0.0% |

### 10 9900% Prizes

| | $\alpha$ = 1.04, b = 150 | | | $\alpha$ = 1.12, b = 250 | | |
|---|---|---|---|---|---|---|
| **Risk** | **Random** | **Bracket** | **Percentage** | **Random** | **Bracket** | **Percentage** |
| **5%** | 1.907876 | 1.714167 | 1.0% | 1.99847 | 1.796518 | 1.5% |
| **1%** | 7.602774 | 7.3186 | 35.5% | 6.912167 | 6.644698 | 36.5% |
| **0.1%** | 62.95666 | 64.64049 | 48.5% | 54.43648 | 54.97797 | 52.0% |
| **0.01%** | 943.7768 | 942.6159 | 33.5% | 739.5197 | 770.5159 | 34.5% |
| **Worst** | 2844.366 | 1382.964 | 0.0% | 2166.284 | 1046.519 | 0.0% |

## Table 3

Difference in the payout average after bracketing. Positive percentages reflect the random prizes having a higher average payoff and negative percentages reflect that the bracketing average was higher.

| Risk | $\alpha$ = 1.04, b = 150 | | | | $\alpha$ = 1.12, b = 250 | | | |
|---|---|---|---|---|---|---|---|---|
| | 100% | 200% | 900% | 9900% | 100% | 200% | 900% | 9900% |
| 5% | 9.7% | 10.0% | 12.3% | 11.3% | 8.5% | 9.9% | 12.1% | 11.2% |
| 1% | 20.3% | 7.7% | 5.0% | 3.9% | 8.6% | 6.0% | 3.3% | 4.0% |
| 0.1% | 5.8% | 1.6% | -10.8% | -2.6% | 6.0% | 1.9% | 5.1% | -1.0% |
| 0.01% | 28.8% | 16.0% | 2.3% | 0.1% | 28.3% | 16.1% | 2.7% | -4.0% |
| VAR Average | 16.1% | 8.8% | 2.2% | 3.2% | 12.9% | 8.5% | 5.8% | 2.6% |
| Worst | 322.1% | 294.2% | 216.1% | 105.7% | 362.5% | 327.6% | 231.9% | 107.0% |

## Table 4

Standard Deviation of the payouts using $\alpha$ = 1.04, b = 150

| Risk | Random | | | | | Bracket | | | | |
|---|---|---|---|---|---|---|---|---|---|---|
| | 100% | 200% | 900% | 9900% | Mean | 100% | 200% | 900% | 9900% | Mean |
| 5% | 0.5 | 0.5 | 0.4 | 0.4 | 0.5 | 0.5 | 0.5 | 0.4 | 0.4 | 0.4 |
| 1% | 12.0 | 4.2 | 2.1 | 1.8 | 5.0 | 11.3 | 3.5 | 2.1 | 1.8 | 4.7 |
| 0.1% | 14.2 | 28.8 | 84.3 | 26.4 | 38.4 | 14.4 | 28.9 | 106.0 | 28.6 | 44.5 |
| 0.01% | 14.8 | 29.1 | 144.5 | 1183.8 | 343.0 | 14.4 | 28.9 | 145.0 | 1167.3 | 338.9 |
| Mean | 10.4 | 15.6 | 57.8 | 303.1 | | 10.2 | 15.5 | 63.3 | 299.5 | |

Standard Deviation of the payouts using $\alpha$ = 1.12, b = 250

| Risk | Random | | | | | Bracket | | | | |
|---|---|---|---|---|---|---|---|---|---|---|
| | 100% | 200% | 900% | 9900% | Mean | 100% | 200% | 900% | 9900% | Mean |
| 5% | 0.4 | 0.4 | 0.4 | 0.4 | 0.4 | 0.4 | 0.3 | 0.3 | 0.3 | 0.3 |
| 1% | 10.6 | 3.9 | 1.5 | 1.4 | 4.3 | 9.9 | 4.1 | 1.6 | 1.3 | 4.2 |
| 0.1% | 13.2 | 26.7 | 86.8 | 25.6 | 38.1 | 13.3 | 26.8 | 77.1 | 23.5 | 35.2 |
| 0.01% | 13.9 | 26.8 | 134.1 | 987.0 | 290.4 | 13.3 | 26.8 | 134.3 | 1067.3 | 310.4 |
| Mean | 9.5 | 14.4 | 55.7 | 253.6 | | 9.2 | 14.5 | 53.3 | 273.1 | |

## Table 5

The columns that begin with "Uncapped" and cap numbers represent the average of 2000 payouts relative to the expected payout for that sample. The percentages represent the percentage of payouts that were higher relative to the expected payout after the cap was lowered (e.g. 250,000 is compared to Uncapped, 10,000 is compared to 50,000)

1000 100% Prizes, $\alpha$ = 1.04, b = 150

| Risk | Uncapped | 250,000 | Percentage | 50,000 | Percentage | 10,000 | Percentage |
|---|---|---|---|---|---|---|---|
| 5% | 2.014 | 1.391 | 10.15% | 1.199 | 0.00% | 1.105 | 0.00% |
| 1% | 9.932 | 1.599 | 0.90% | 1.293 | 0.00% | 1.150 | 0.00% |
| 0.1% | 17.205 | 1.599 | 0.00% | 1.293 | 0.00% | 1.150 | 0.00% |
| Worst | 62.012 | 43.093 | 0.00% | 31.744 | 0.00% | 13.701 | 0.00% |

### 1000 100% Prizes, α = 1.12, b = 250

| Risk | Uncapped | 250,000 | Percentage | 50,000 | Percentage | 10,000 | Percentage |
|---|---|---|---|---|---|---|---|
| **5%** | 1.790 | 1.311 | 6.55% | 1.161 | 0.00% | 1.085 | 0.00% |
| **1%** | 6.844 | 1.475 | 0.60% | 1.236 | 0.00% | 1.121 | 0.00% |
| **0.1%** | 11.873 | 1.475 | 0.00% | 1.236 | 0.00% | 1.121 | 0.00% |
| **Worst** | 50.887 | 36.131 | 0.00% | 26.119 | 0.00% | 10.050 | 0.00% |

### 500 200% Prizes, α = 1.04, b = 150

| Risk | Uncapped | 250,000 | Percentage | 50,000 | Percentage | 10,000 | Percentage |
|---|---|---|---|---|---|---|---|
| **5%** | 1.991 | 1.579 | 16.90% | 1.289 | 0.00% | 1.150 | 0.00% |
| **1%** | 7.699 | 1.908 | 1.95% | 1.430 | 0.00% | 1.217 | 0.00% |
| **0.1%** | 7.699 | 1.908 | 1.95% | 1.430 | 0.00% | 1.217 | 0.00% |
| **Worst** | 113.713 | 70.733 | 0.00% | 44.950 | 0.00% | 13.701 | 0.00% |

### 500 200% Prizes, α = 1.12, b = 250

| Risk | Uncapped | 250,000 | Percentage | 50,000 | Percentage | 10,000 | Percentage |
|---|---|---|---|---|---|---|---|
| **5%** | 1.810 | 1.460 | 10.95% | 1.233 | 0.00% | 1.122 | 0.00% |
| **1%** | 5.698 | 1.718 | 1.30% | 1.346 | 0.00% | 1.175 | 0.00% |
| **0.1%** | 5.698 | 1.718 | 1.30% | 1.346 | 0.00% | 1.175 | 0.00% |
| **Worst** | 90.757 | 57.931 | 0.00% | 35.658 | 0.00% | 10.050 | 0.00% |

### 100 900% Prizes, α = 1.04, b = 150

| Risk | Uncapped | 250,000 | Percentage | 50,000 | Percentage | 10,000 | Percentage |
|---|---|---|---|---|---|---|---|
| **5%** | 1.922 | 2.679 | 98.25% | 1.700 | 0.00% | 1.349 | 0.00% |
| **1%** | 7.138 | 3.430 | 6.45% | 2.095 | 0.00% | 1.519 | 0.00% |
| **0.1%** | 7.138 | 3.430 | 6.45% | 2.095 | 0.00% | 1.519 | 0.00% |
| **Worst** | 453.945 | 181.762 | 0.00% | 53.908 | 0.00% | 13.701 | 0.00% |

### 100 900% Prizes, α = 1.12, b = 250

| Risk | Uncapped | 250,000 | Percentage | 50,000 | Percentage | 10,000 | Percentage |
|---|---|---|---|---|---|---|---|
| **5%** | 1.896 | 2.305 | 94.75% | 1.561 | 0.00% | 1.282 | 0.00% |
| **1%** | 5.853 | 2.928 | 3.75% | 1.875 | 0.00% | 1.416 | 0.00% |
| **0.1%** | 5.853 | 2.928 | 3.75% | 1.875 | 0.00% | 1.416 | 0.00% |
| **Worst** | 340.707 | 142.485 | 0.00% | 40.654 | 0.00% | 10.050 | 0.00% |

### 10 9900% Prize, α = 1.04, b = 150

| Risk | Uncapped | 250,000 | Percentage | 50,000 | Percentage | 10,000 | Percentage |
|---|---|---|---|---|---|---|---|
| **5%** | 1.847 | 2.771 | 100.00% | 3.324 | 100.00% | 2.218 | 0.00% |
| **1%** | 7.412 | 11.098 | 100.00% | 6.053 | 0.90% | 3.067 | 0.00% |
| **0.1%** | 7.412 | 11.098 | 100.00% | 6.053 | 0.90% | 3.067 | 0.00% |
| **Worst** | 3007.150 | 224.728 | 0.00% | 53.908 | 0.00% | 13.701 | 0.00% |

10 9900% Prize, α = 1.12, b = 250

| Risk | Uncapped | 250,000 | Percentage | 50,000 | Percentage | 10,000 | Percentage |
|---|---|---|---|---|---|---|---|
| 5% | 2.019 | 2.625 | 100.00% | 3.037 | 100.00% | 1.982 | 0.00% |
| 1% | 7.031 | 9.117 | 99.65% | 4.855 | 0.15% | 2.587 | 0.00% |
| 0.1% | 7.031 | 9.117 | 99.65% | 4.855 | 0.15% | 2.587 | 0.00% |
| Worst | 2090.488 | 175.728 | 0.00% | 40.654 | 0.00% | 10.050 | 0.00% |